\documentclass[]{aa}

\def\rmit#1{{\it #1}}              
\def\specchar#1{{\sc #1}}
\def\FeI{\mbox{Fe\,\specchar{i}}}

\def\CaII{\mbox{Ca\,\specchar{ii}}}

\def\eg{\rmit{e.g.}}

\usepackage{natbib}
\bibpunct{(}{)}{;}{a}{}{,} 
\usepackage{graphicx}
\usepackage{amsmath}

\usepackage{color}

\usepackage{hyperref}
\hypersetup{
    final=true,
    pageanchor=true,
    colorlinks=true,
    breaklinks=true,
    linkcolor=blue,
    citecolor=blue,
    urlcolor=blue,
    pdfpagemode=UseNone,
    pdftitle={Dependence of sunspot photospheric waves on the depth of the source of solar $p-$modes},
    pdfauthor={T. Felipe et al.},
    pdfsubject={Solar Physics},
    pdfkeywords={Methods: numerical -- Sun: oscillations -- Sun: helioseismology -- Sun: photosphere -- sunspots}}

\titlerunning{Photospheric waves and source depths}

\authorrunning{Felipe \& Khomenko}

\begin{document}

\title{Dependence of sunspot photospheric waves on the depth of the source of solar $p-$modes}

\author{T. Felipe\inst{\ref{inst1},\ref{inst2}}
\and E. Khomenko\inst{\ref{inst1},\ref{inst2}}
}


\institute{Instituto de Astrof\'{\i}sica de Canarias, 38205, C/ V\'{\i}a L{\'a}ctea, s/n, La Laguna, Tenerife, Spain\label{inst1}
\and 
Departamento de Astrof\'{\i}sica, Universidad de La Laguna, 38205, La Laguna, Tenerife, Spain\label{inst2}
}

\abstract{Photospheric waves in sunspots moving radially outwards at speeds faster than the characteristic wave velocities have been recently detected. It has been suggested that they are the visual pattern of $p-$modes excited around 5 Mm beneath the sunspot. Using numerical simulations, we have performed a parametric study of the waves observed at the photosphere and higher layers produced by sources located at different depths. The observational measurements are consistent with waves driven between approximately 1 Mm and 5 Mm below the sunspot surface.  }


\maketitle


\section{Introduction}
\vspace{-0.2cm}

Sunspots' atmospheres exhibit a wide variety of oscillatory phenomena. Their photosphere is dominated by five-minute oscillations  \citep{Thomas+etal1982}, while at the chromosphere three-minutes oscillations become stronger. These chromospheric waves have been identified as upward propagating, field aligned, slow-mode waves \citep{Centeno+etal2006, Bloomfield+etal2007, Felipe+etal2010b}, which can develop into shocks and lead to the appearance of umbral flashes \citep{RouppevanderVoort+etal2003,delaCruzRodriguez+etal2013}. Umbral flashes were first detected as sudden brightness increases in the core of \CaII\ lines \citep{Beckers+Tallant1969}. They are found to be related to running penumbral waves \citep[RPWs][]{Zirin+Stein1972,Giovanelli1972}, which are observed as disturbances propagating radially outwards at the chromospheric penumbra. The origin of RPWs has been a matter of debate over the years. Currently it is acknowledged that they are the visual pattern of slow low-$\beta$ plasma waves propagating up from the photosphere along the inclined magnetic field lines \citep{Bloomfield+etal2007, Jess+etal2013, Madsen+etal2015}. According to the current understanding, all the sunspot waves are different manifestations of a common process of magnetoacoustic wave propagation \citep{Khomenko+Collados2015}. 

This picture has recently been reinforced by the report of the photospheric counterpart of RPWs \citep{LohnerBottcher+BelloGonzalez2015}. It was observed that, due to the variation of the inclination of the penumbral magnetic field with height, the apparent horizontal velocity of the RPWs decreases from the photosphere to the chromosphere. In addition to this direct observation, a similar photospheric wave has been detected using helioseismic techniques \citep{Zhao+etal2015}. The phase velocity of this wave (around \hbox{45 km s$^{-1}$}) was found to be faster than that of the MHD waves at the photosphere. It has been proposed that this fast-moving wave corresponds to the visual pattern of $p-$modes driven beneath the sunspot. 

In this Letter, we aim to explore the nature of RPWs by analyzing numerical simulations of wave propagation in sunspots excited by sources at different depths. Under the assumption that the photospheric RPWs and the fast-moving waves are the same phenomenon, we will use indistinctly both concepts in order to refer to these waves. We describe the methods used for the development of the numerical simulations in \hbox{Sect. \ref{sect:simulations}}, the results are presented in \hbox{Sect. \ref{sect:results}}, and the conclusions are discussed in \hbox{Sect. \ref{sect:conclusions}}.

\vspace{-0.4cm}

\section{Numerical simulations}
\label{sect:simulations}

We have used the code MANCHA \citep{Khomenko+Collados2006, Felipe+etal2010a} for solving the ideal MHD equations for perturbations, which are obtained after explicitly removing the equilibrium state from the continuity, momentum, internal energy, and induction equations. The numerical simulations have been restricted to the linear regime by choosing a small amplitude for the driver. Periodic boundary conditions were imposed at the horizontal directions and Perfect Matched Layers \citep{Berenger1996}, which damp waves producing minimum reflection, were used at the top and bottom boundaries. 

Simulations have been computed under the 2.5D approximation. This means that vectors keep their three components, but the derivatives in the $Y$ direction have been neglected. Photospheric RPWs have been reported using time-slice analyses in the radial direction of the sunspot \citep{LohnerBottcher+BelloGonzalez2015} or using azimuthal averaged time-distance diagrams \citep{Zhao+etal2015}. In both cases, a two dimensional computation domain crossing the center of the sunspot is enough for comparing the numerical results with the observations. The Alfv\'en speed has been limited to a maximum value of 1000 km s$^{-1}$ by modifying the Lorentz force following \citet{Rempel+etal2009}. According to \citep{Moradi+Cally2014}, this limiter can affect the results if the maximum allowed Alfv\'en speed is not sufficiently higher than the horizontal phase speeds of interest. The chosen value is well above the phase velocities under study.

The sunspot model has been constructed following \citet{Khomenko+Collados2008}. The procedures have been modified in order to extend the atmosphere up to the corona. This method requires the selection of the stratification at the sunspot axis and at a quiet Sun region as boundary conditions. For the photospheric and chromospheric layers, we have chosen VAL-C model \citep{Vernazza+etal1981} for the quiet Sun and the semi-empirical model of \citet{Avrett1981} for the sunspot umbra. For both models, the interior was taken from the convectively stabilized CSM\_B model \citep{Schunker+etal2011} and an isothermal corona with a temperature of one million Kelvin was added following \citet{Santamaria+etal2015}. The photospheric magnetic field at the umbra is 900 G and the Wilson depression was set to \hbox{450 km}. The field strength is significantly lower than that expected in large sunspots, but the model qualitatively reproduces properties of wave propagation through magnetic structures. The computational domain spans from \hbox{$z=-7.30$ Mm} to \hbox{$z=10.65$ Mm}, and it is sampled by 360 equally spaced points in the vertical direction (vertical sampling of 50 km). The vertical position $z=0$ is located at the umbral height where the optical depth at 500 nm ($\tau$) is unity. The horizontal domain covers \hbox{120 Mm} with a spatial sampling of \hbox{150 km}. The axis of the sunspot is located at the center of the domain.

The wave field is driven by spatially localized sources of the vertical force added to the equations. The shape and temporal variation of each source is described by \citet{Parchevsky+etal2008}. This driver generates a wave field that resembles the solar spectrum \citep[\eg,][]{Parchevsky+etal2008, Felipe+etal2016a}. We have performed several independent numerical experiments which differ in the amount and location of the sources. In the first set of simulations, we have computed six simulations with one single source located at the axis of the sunspot but at different depths between $z=-5.3$ Mm and $z=-0.3$ Mm with 1 Mm interval. The sources are placed beneath the sunspot following \citet{Zhao+etal2015} suggestion. In the second set of simulations, we have reproduced the stochastic solar wave field by introducing a new source every 0.5 s at a randomly chosen horizontal location, but with the depth of the sources fixed for each run to the same geometrical depths used for the single source simulations. Each of the sources has the same frequency spectrum used for the single source cases, with a central frequency at 3.3 mHz.

\vspace{-0.4cm}
\section{Results}
\label{sect:results}

Maps of vertical velocity at constant optical depth have been constructed by interpolating at each horizontal position and time step the vertical variation of the $z$ velocity to the height where $\log\tau$ takes the chosen value. For most of the analyses in this Letter we will focus on the maps at $\log\tau=-2$, which provides a good characterization of the formation height of the \FeI\ 630.15 nm line used for the detection of RPWs in actual observations.

\begin{figure}[!ht] 
 \centering
 \includegraphics[width=9cm]{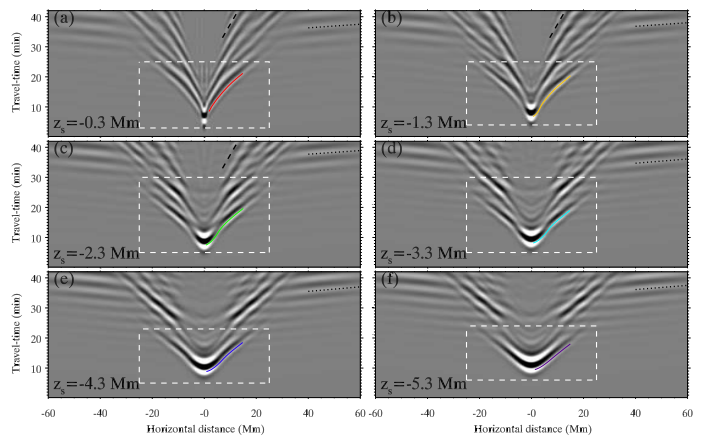}
  \caption{Time-distance diagrams at $\log\tau=-2$ for the simulations with waves driven by a single source located at the center of the sunspot. Each panel corresponds to a different source depth: $z_s=-0.3$ Mm (a),  $z_s=-1.3$ Mm (b), $z_s=-2.3$ Mm (c), $z_s=-3.3$ Mm (d), $z_s=-4.3$ Mm (e), and $z_s=-5.3$ Mm (f). Color lines illustrate the track of the fast-moving waves. Dotted lines in the top right part of all panels show a linear fitting of the reflected fast wave. Dashed lines in panels (a), (b), and (c) show a linear fitting of the helioseismic waves. Dashed white boxes delimite the first wavefronts of the fast-moving wave.}
  \label{fig:TD_1source}  
\end{figure}

\begin{figure}[!ht] 
 \centering
 \includegraphics[width=9cm]{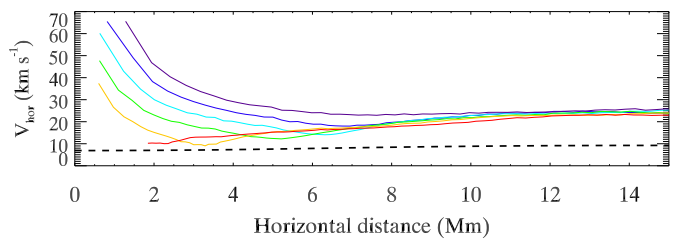}
  \caption{Phase velocity of the fast-moving waves produced by a single source at $z_s=-0.3$ Mm (red), $z_s=-1.3$ Mm (yellow), $z_s=-2.3$ Mm (green), $z_s=-3.3$ Mm (light blue), $z_s=-4.3$ Mm (dark blue), and $z_s=-5.3$ Mm (violet). The dash line shows the fast-wave speed at $\log\tau=-2$.}
  \label{fig:velocity_1source}
\end{figure}

\vspace{-0.4cm}
\subsection{Single source}

Figure \ref{fig:TD_1source} shows the time-distance diagrams of the simulations with one isolated source located at the center of the sunspot at different depths. In all cases the white dashed boxes delimit the first wavefronts of a wave that sweeps outwards with a speed higher than the local fast magnetoacoustic speed (around 10 km s$^{-1}$). These fast-moving waves are produced because the wavefronts driven in the interior of the Sun need different time to reach the photosphere at different horizontal positions. The sources mainly generate fast high-$\beta$ plasma acoustic waves with approximately spherical wavefronts. In the interior, they propagate in all directions at the local sound speed. The region of the wavefront above the source is the first one to become visible at the photosphere. It is then followed by the surrounding parts of the wavefront. As the wavefronts reach the photosphere at progressively farther horizontal locations, an apparent radially propagating wave appears at $\log\tau=-2$.       

The color lines in Figure \ref{fig:TD_1source} illustrate the position of the wavefront of the fast-moving wave. The phase velocity of this apparent wave is given in \hbox{Fig. \ref{fig:velocity_1source}}. When the first appearance of the wave is produced, the wavefront propagating upward is nearly parallel to the $\log\tau=-2$ surface. The initial velocity of the fast-moving wave is very high since regions of the wavefront at a certain horizontal distance reach the photosphere almost simultaneously. For deeper sources the wavefront of the incident wave is flatter and, thus, the fast-moving wave initially shows a higher apparent velocity. The apparent velocity is reduced with the distance. At horizontal distances around \hbox{12 Mm} from the origin, the velocity of the fast-moving waves is around \hbox{25 km s$^{-1}$} for all the simulations with different source depths. The dashed line in Fig. \ref{fig:velocity_1source} indicates the velocity of the fast wave at the surface $\log\tau=-2$. It is between 7 and \hbox{10 km s$^{-1}$} for all horizontal locations, which is clearly slower than the phase velocity of the apparent wave. The fast wave speed for a sunspot model with a more realistic photospheric magnetic field strength around 3000 G is below \hbox{15 km s$^{-1}$}, still lower than the measured phase velocity.

Other wave branches are present in our simulations. The three simulations with the shallower sources (panels a, b, and c from Fig. \ref{fig:TD_1source}) show the usual helioseismic waves after about 23 minutes of simulation. In each of these panels, one of those wavefronts is indicated by a dashed line obtained from a linear fit. In the three cases it corresponds to a phase velocity of \hbox{10 km s$^{-1}$}, in agreement with the local fast velocity. The helioseismic wave is barely visible in the cases with the source located at depths below \hbox{$z=-2.3$ Mm}.

Another distinct branch with much higher phase velocity is found at longer radial distances in all simulations. It appears around 10 minutes after the first velocity perturbations is visible, but it is more easily seen at later time steps. We will refer to them as super-fast-moving waves. One of those wavefronts is marked in all panels of Fig. \ref{fig:TD_1source} by a dotted line. The phase velocity of super-fast moving waves is between 222 and \hbox{277 km s$^{-1}$}. A careful inspection of the temporal evolution of the simulations reveals that these waves are the visual pattern of the fast magnetic wave coming after reflection from the upper part of the atmosphere. The source (located in the interior in a high-$\beta$ region) drives fast acoustic waves which reach the photosphere. At the surface $\beta\approx1$ they are converted into fast magnetic and slow acoustic waves (low-$\beta$ region). The fast magnetic wave is refracted due to the gradients of the Alfv\'en speed \citep{Rosenthal+etal2002, Khomenko+Collados2006} and they return toward the interior of the Sun. In our simulations, fast waves are detected up to the high-chromosphere, where only almost-vertical fast waves can penetrate. Waves with higher inclinations are reflected at lower heights around the mid-chromosphere. Their wavefronts form a certain angle with respect to the surface $\log\tau=-2$, and the section closer to the umbra reach the photosphere earlier than the sections at longer radial distances. They appear as an apparent wave propagating radially outwards. This effect is similar to that producing the fast-moving wave, but the later is generated by $p-$modes coming from the interior and the former is due to the return of atmospheric waves. Independent simulations (not shown in the figures) with the same configuration but with the magnetic field set to zero have been computed. In this case the super-fast-moving waves do not appear, confirming that their presence is associated with mode conversion. Since the Lorentz force has been limited to a maximum value of the Alfv\'en speed of \hbox{1000 km s$^{-1}$}, these simulations do not capture all the physics of the fast wave refraction. However, we expect that the results found in our simulations qualitatively match the unrestricted process.

\vspace{-0.4cm}
\subsection{Stochastic sources}

The cross-correlation of the oscillatory signal measured at one location with those observed at other locations is a common helioseismic approach for the analysis and interpretation of stochastic wavefields \citep[\eg,][]{Duvall+etal1993, Cameron+etal2008,Zhao+etal2011}. For each map at $\log\tau$, we have cross-correlated the vertical velocity signal at \hbox{$x_{\rm ref}=7.5$ Mm} with the vertical velocity at all other locations inside the area of interest. The location $x_{\rm ref}$ corresponds to the penumbra, at approximately the same radial distance where the RPWs were reported by \citet{LohnerBottcher+BelloGonzalez2015}. The cross-correlation at time lag $t$ indicates the position of a wave packet at time $t$ after (before) it has left (arrived) to $x_{\rm ref}$.

The cross-correlations of maps at a height given by different optical depths (between the base of the photosphere and the low chromosphere) reveals that the aparent velocity of the wave moving radially outwards monotonically decreases with height. The variation of the velocity with height shows a similar trend for all the simulations, independently of the depth of the sources. This reduction of the apparent horizontal velocity of RPWs from the photosphere to the chromosphere is in agreement with the recent observations from \citet{LohnerBottcher+BelloGonzalez2015}.

\begin{figure}[!ht] 
 \centering
 \includegraphics[width=9cm]{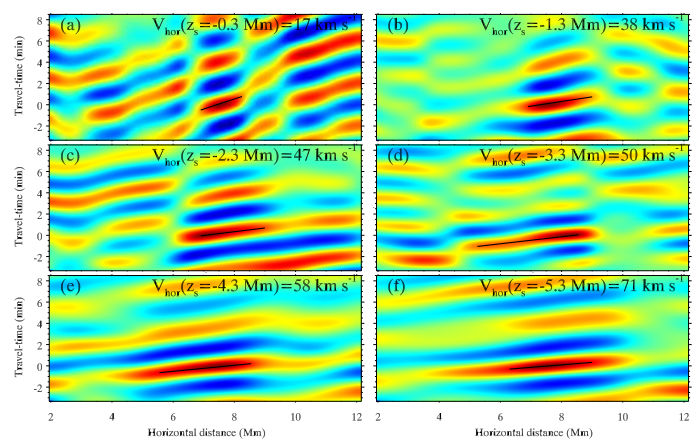}
  \caption{Time-distance diagrams at $log\tau =-2$ for the simulations with stochastic driver obtained from the cross-correlation between the vertical velocity at \hbox{$x=7.5$ Mm} and the vertical velocity at the $x$ positions indicated in the horizontal axis. Each panel corresponds to a different source depth: $z_s=-0.3$ Mm (a),  \hbox{$z_s=-1.3$ Mm (b)}, \hbox{$z_s=-2.3$ Mm (c)}, \hbox{$z_s=-3.3$ Mm (d)}, \hbox{$z_s=-4.3$ Mm (e)}, and \hbox{$z_s=-5.3$ Mm (f)}.}
  \label{fig:cross_correlations_tau-2}
\end{figure}

We found that the phase velocity of the wave detected in the cross-correlation does depend on the depth of the stochastic sources. Figure \ref{fig:cross_correlations_tau-2} illustrates the time-distance diagrams at $\log\tau=-2$ for all the simulations. For the cases with \hbox{$z_s=-1.3$ Mm} and deeper, the cross-correlation clearly isolates the wavefront of a wave moving from the inner to the outer penumbra. In the case of the simulation with the sources at $z_s=-0.3$ Mm, high correlation and anti-correlation is found for several wavefronts, but the radially outward movement is also prominent. Deeper sources lead to faster horizontal velocities, similar to the results found for the single source simulations. A maximum horizontal velocity at $\log\tau=-2$ of \hbox{71 km s$^{-1}$} is obtained when the sources are located at \hbox{$z_s=-5.3$ Mm}, while shallow sources at \hbox{$z_s=-0.3$ Mm} produce a horizontal velocity of \hbox{17 km s$^{-1}$}. The cross-correlation of an independent simulation with no sources located beneath the sunspot within radial distances of 10 Mm shows a radially inward propagating wave, excited by the sources located out of the spot.

\vspace{-0.5cm}

\section{Discussion and conclusions}
\label{sect:conclusions}

\citet{LohnerBottcher+BelloGonzalez2015} recently reported the observation of RPWs at photospheric layers. They present apparent horizontal velocities of \hbox{$51\pm 13$ km s$^{-1}$}, which decrease to \hbox{$37\pm 10$ km s$^{-1}$} at the chromosphere. A similar photospheric wave was independently detected through helioseismic analyses by \citet{Zhao+etal2015}. Since the apparent speed of the wave is faster than magnetoacoustic or Alfv\'en waves speed at the photosphere of sunspots, \citet{Zhao+etal2015} proposed that the observed waves are the wavefronts of acoustic waves driven about 5 Mm beneath sunspot's surface, which sweeps across the photophere. 
 
Motivated by this hypothesis, we have performed a parametric study of the dependence of the visible waves at the photosphere and higher layers with the depth of the sources driving them. Twelve numerical simulations have been developed, six of them with a single source and the other six with stochastic sources. In both sets of simulations the drivers are located at different depths between $-5.3$ and \hbox{$-0.3$ Mm}. The results show a prominent wave propagating radially outwards at a faster velocity than that of the MHD waves at the photosphere of sunspots. This fast-moving wave is produced by the $p-$modes generated in the interior by the sources, whose almost spherical wavefronts expand and reach the surface at different times at each horizontal location. A simulation similar to those with a single source has recently proved that this model qualitatively resembles the observed properties and the shape of the atmospheric wavefronts (from the photosphere to the corona) retrieved from an helioseismic analysis \citep{Zhao+etal2016}. 

Convective motions near the surface of the Sun are known to be the source of excitation of the $p-$modes. Many works have attempted to infer the properties of these sources from the analysis of the velocity and intensity power spectra asymmetries \citep[\eg,][]{Roxburgh+Vorontsov1997, Chaplin+Appourchaux1999, Kumar+Basu1999, Nigam+Kosovichev1999}. The depth of the sources obtained from these analyses is highly sensitive to the model selection, and these works have provided estimates between 75 and 1500 km beneath the photosphere. In addition, more recent works have found that the acoustic source depth cannot be uniquely determined from the properties of the resonant $p$-modes \citep{Jefferies+etal2003,Wachter+Kosovichev2005}. From a comparison between the observed photospheric RPWs velocities and the source-depth dependence of the apparent wave velocity in the simulations, we can estimate the location of the sources in sunspots. The velocity of the fast-moving wave obtained from the cross-correlation of the simulations with stochastic sources is within the limits of the speed of the photospheric RPWs measured by \citet{LohnerBottcher+BelloGonzalez2015} for source depths between \hbox{$z_s=-1.3$ Mm} and \hbox{$z_s=-4.3$ Mm}. As reported by \citet{Zhao+etal2015}, the fast-moving wave only appears in sunspots. Our results suggest that in sunspots, the source of excitation of $p-$modes is extended to deeper regions that can be as deep as almost 5 Mm beneath the surface.

The propagation velocity of the fast-moving wave strongly depends on the distance from the horizontal position of the source, as seen in Fig. \ref{fig:velocity_1source}. When this distance is shorter, the apparent velocity is maximum and it increases with the depth of the source. At larger horizontal distances all the source depths produce similar apparent velocity. The velocities of the fast-moving wave measured from the cross-correlation in the stochastic simulations is in good agreement with those obtained from single sources at short horizontal distances from the sources, confirming that the wave detected using helioseismic methods corresponds to the same fast-moving wave identified in the single source simulations. Our simulations also show that sources deeper than $z_s=-2.3$ Mm are not efficient driving the usual helioseismic waves. Only the simulations with shallower sources lead to the appearance of wave propagating at the fast magnetoacoustic speed (panels a, b, and c from Fig. \ref{fig:TD_1source}). 

The simulations show another wave that we have called super-fast-moving wave. This is also an apparent wave, but produced by the fast magnetic waves which are refracted at atmospheric layers. The study of this wave can potentially provide some insights about the magnetic field in the solar atmosphere, although its detection can be challenging due to its low amplitude in comparison with the photospheric RPWs and the $p-$modes and its fast apparent speed.








 

\begin{acknowledgements} 

We acknowledge the financial support by the Spanish Ministry of Economy
and Competitiveness (MINECO) through projects AYA2014-55078-P, AYA2014-60476-P, and AYA2014-60833-P. This work used Teide High-
Performance Computing facilities at Instituto Tecnol\'ogico y de Energ\'ias Renovables (ITER, SA)
and MareNostrum supercomputer at Barcelona Supercomputing Center.

\end{acknowledgements}

\vspace{-1cm}

\bibliographystyle{aa} 
\bibliography{biblio.bib}

\begin{thebibliography}{37}
\expandafter\ifx\csname natexlab\endcsname\relax\def\natexlab#1{#1}\fi

\bibitem[{{Avrett}(1981)}]{Avrett1981}
{Avrett}, E.~H. 1981, in The Physics of Sunspots, ed. L.~E. {Cram} \& J.~H.
  {Thomas}, 235--255

\bibitem[{{Beckers} \& {Tallant}(1969)}]{Beckers+Tallant1969}
{Beckers}, J.~M. \& {Tallant}, P.~E. 1969, \solphys, 7, 351

\bibitem[{{Berenger}(1996)}]{Berenger1996}
{Berenger}, J.~P. 1996, Journal of Computational Physics, 127, 363

\bibitem[{{Bloomfield} {et~al.}(2007){Bloomfield}, {Solanki}, {Lagg},
  {Borrero}, \& {Cally}}]{Bloomfield+etal2007}
{Bloomfield}, D.~S., {Solanki}, S.~K., {Lagg}, A., {Borrero}, J.~M., \&
  {Cally}, P.~S. 2007, \aap, 469, 1155

\bibitem[{{Cameron} {et~al.}(2008){Cameron}, {Gizon}, \&
  {Duvall}}]{Cameron+etal2008}
{Cameron}, R., {Gizon}, L., \& {Duvall}, Jr., T.~L. 2008, \solphys, 251, 291

\bibitem[{{Centeno} {et~al.}(2006){Centeno}, {Collados}, \& {Trujillo
  Bueno}}]{Centeno+etal2006}
{Centeno}, R., {Collados}, M., \& {Trujillo Bueno}, J. 2006, \apj, 640, 1153

\bibitem[{{Chaplin} \& {Appourchaux}(1999)}]{Chaplin+Appourchaux1999}
{Chaplin}, W.~J. \& {Appourchaux}, T. 1999, \mnras, 309, 761

\bibitem[{{de la Cruz Rodr{\'{\i}}guez} {et~al.}(2013){de la Cruz
  Rodr{\'{\i}}guez}, {Rouppe van der Voort}, {Socas-Navarro}, \& {van
  Noort}}]{delaCruzRodriguez+etal2013}
{de la Cruz Rodr{\'{\i}}guez}, J., {Rouppe van der Voort}, L., {Socas-Navarro},
  H., \& {van Noort}, M. 2013, \aap, 556, A115

\bibitem[{{Duvall} {et~al.}(1993){Duvall}, {Jefferies}, {Harvey}, \&
  {Pomerantz}}]{Duvall+etal1993}
{Duvall}, Jr., T.~L., {Jefferies}, S.~M., {Harvey}, J.~W., \& {Pomerantz},
  M.~A. 1993, \nat, 362, 430

\bibitem[{{Felipe} {et~al.}(2016){Felipe}, {Braun}, {Crouch}, \&
  {Birch}}]{Felipe+etal2016a}
{Felipe}, T., {Braun}, D.~C., {Crouch}, A.~D., \& {Birch}, A.~C. 2016, \apj,
  829, 67

\bibitem[{{Felipe} {et~al.}(2010{\natexlab{a}}){Felipe}, {Khomenko}, \&
  {Collados}}]{Felipe+etal2010a}
{Felipe}, T., {Khomenko}, E., \& {Collados}, M. 2010{\natexlab{a}}, \apj, 719,
  357

\bibitem[{{Felipe} {et~al.}(2010{\natexlab{b}}){Felipe}, {Khomenko},
  {Collados}, \& {Beck}}]{Felipe+etal2010b}
{Felipe}, T., {Khomenko}, E., {Collados}, M., \& {Beck}, C. 2010{\natexlab{b}},
  Apj, 722, 131

\bibitem[{{Giovanelli}(1972)}]{Giovanelli1972}
{Giovanelli}, R.~G. 1972, \solphys, 27, 71

\bibitem[{{Jefferies} {et~al.}(2003){Jefferies}, {Severino}, {Moretti},
  {Oliviero}, \& {Giebink}}]{Jefferies+etal2003}
{Jefferies}, S.~M., {Severino}, G., {Moretti}, P.-F., {Oliviero}, M., \&
  {Giebink}, C. 2003, \apjl, 596, L117

\bibitem[{{Jess} {et~al.}(2013){Jess}, {Reznikova}, {Van Doorsselaere}, {Keys},
  \& {Mackay}}]{Jess+etal2013}
{Jess}, D.~B., {Reznikova}, V.~E., {Van Doorsselaere}, T., {Keys}, P.~H., \&
  {Mackay}, D.~H. 2013, \apj, 779, 168

\bibitem[{{Khomenko} \& {Collados}(2006)}]{Khomenko+Collados2006}
{Khomenko}, E. \& {Collados}, M. 2006, \apj, 653, 739

\bibitem[{{Khomenko} \& {Collados}(2008)}]{Khomenko+Collados2008}
{Khomenko}, E. \& {Collados}, M. 2008, \apj, 689, 1379

\bibitem[{{Khomenko} \& {Collados}(2015)}]{Khomenko+Collados2015}
{Khomenko}, E. \& {Collados}, M. 2015, Living Reviews in Solar Physics, 12, 6

\bibitem[{{Kumar} \& {Basu}(1999)}]{Kumar+Basu1999}
{Kumar}, P. \& {Basu}, S. 1999, \apj, 519, 396

\bibitem[{{L{\"o}hner-B{\"o}ttcher} \& {Bello
  Gonz{\'a}lez}(2015)}]{LohnerBottcher+BelloGonzalez2015}
{L{\"o}hner-B{\"o}ttcher}, J. \& {Bello Gonz{\'a}lez}, N. 2015, \aap, 580, A53

\bibitem[{{Madsen} {et~al.}(2015){Madsen}, {Tian}, \&
  {DeLuca}}]{Madsen+etal2015}
{Madsen}, C.~A., {Tian}, H., \& {DeLuca}, E.~E. 2015, \apj, 800, 129

\bibitem[{{Moradi} \& {Cally}(2014)}]{Moradi+Cally2014}
{Moradi}, H. \& {Cally}, P.~S. 2014, \apjl, 782, L26

\bibitem[{{Nigam} \& {Kosovichev}(1999)}]{Nigam+Kosovichev1999}
{Nigam}, R. \& {Kosovichev}, A.~G. 1999, \apjl, 514, L53

\bibitem[{{Parchevsky} {et~al.}(2008){Parchevsky}, {Zhao}, \&
  {Kosovichev}}]{Parchevsky+etal2008}
{Parchevsky}, K.~V., {Zhao}, J., \& {Kosovichev}, A.~G. 2008, \apj, 678, 1498

\bibitem[{{Rempel} {et~al.}(2009){Rempel}, {Sch{\"u}ssler}, \&
  {Kn{\"o}lker}}]{Rempel+etal2009}
{Rempel}, M., {Sch{\"u}ssler}, M., \& {Kn{\"o}lker}, M. 2009, \apj, 691, 640

\bibitem[{{Rosenthal} {et~al.}(2002){Rosenthal}, {Bogdan}, {Carlsson}, {Dorch},
  {Hansteen}, {McIntosh}, {McMurry}, {Nordlund}, \&
  {Stein}}]{Rosenthal+etal2002}
{Rosenthal}, C.~S., {Bogdan}, T.~J., {Carlsson}, M., {et~al.} 2002, \apj, 564,
  508

\bibitem[{{Rouppe van der Voort} {et~al.}(2003){Rouppe van der Voort},
  {Rutten}, {S{\"u}tterlin}, {Sloover}, \&
  {Krijger}}]{RouppevanderVoort+etal2003}
{Rouppe van der Voort}, L.~H.~M., {Rutten}, R.~J., {S{\"u}tterlin}, P.,
  {Sloover}, P.~J., \& {Krijger}, J.~M. 2003, \aa, 403, 277

\bibitem[{{Roxburgh} \& {Vorontsov}(1997)}]{Roxburgh+Vorontsov1997}
{Roxburgh}, I.~W. \& {Vorontsov}, S.~V. 1997, \mnras, 292, L33

\bibitem[{{Santamaria} {et~al.}(2015){Santamaria}, {Khomenko}, \&
  {Collados}}]{Santamaria+etal2015}
{Santamaria}, I.~C., {Khomenko}, E., \& {Collados}, M. 2015, \aap, 577, A70

\bibitem[{{Schunker} {et~al.}(2011){Schunker}, {Cameron}, {Gizon}, \&
  {Moradi}}]{Schunker+etal2011}
{Schunker}, H., {Cameron}, R.~H., {Gizon}, L., \& {Moradi}, H. 2011, \solphys,
  271, 1

\bibitem[{{Thomas} {et~al.}(1982){Thomas}, {Cram}, \& {Nye}}]{Thomas+etal1982}
{Thomas}, J.~H., {Cram}, L.~E., \& {Nye}, A.~H. 1982, \nat, 297, 485

\bibitem[{{Vernazza} {et~al.}(1981){Vernazza}, {Avrett}, \&
  {Loeser}}]{Vernazza+etal1981}
{Vernazza}, J.~E., {Avrett}, E.~H., \& {Loeser}, R. 1981, \apjs, 45, 635

\bibitem[{{Wachter} \& {Kosovichev}(2005)}]{Wachter+Kosovichev2005}
{Wachter}, R. \& {Kosovichev}, A.~G. 2005, \apj, 627, 550

\bibitem[{{Zhao} {et~al.}(2015){Zhao}, {Chen}, {Hartlep}, \&
  {Kosovichev}}]{Zhao+etal2015}
{Zhao}, J., {Chen}, R., {Hartlep}, T., \& {Kosovichev}, A.~G. 2015, \apjl, 809,
  L15

\bibitem[{{Zhao} {et~al.}(2016){Zhao}, {Felipe}, {Chen}, \&
  {Khomenko}}]{Zhao+etal2016}
{Zhao}, J., {Felipe}, T., {Chen}, R., \& {Khomenko}, E. 2016, \apjl, 830, L17

\bibitem[{{Zhao} {et~al.}(2011){Zhao}, {Kosovichev}, \&
  {Ilonidis}}]{Zhao+etal2011}
{Zhao}, J., {Kosovichev}, A.~G., \& {Ilonidis}, S. 2011, \solphys, 268, 429

\bibitem[{{Zirin} \& {Stein}(1972)}]{Zirin+Stein1972}
{Zirin}, H. \& {Stein}, A. 1972, \apjl, 178, L85+

\end{thebibliography}

\end{document}